\documentclass[12pt]{iopart}
\usepackage{graphicx}
\begin{document}

\title[Density fluctuations in non--equilibrium
1st order phase transition]
{Density Fluctuations as Signature of 
\\
a Non--Equilibrium
First Order Phase Transition}

\author{C Sasaki$^1$, B Friman$^2$ and K Redlich$^{3,4}$}

\address{$^1$ Physik-Department,
Technische Universit\"{a}t M\"{u}nchen,
D-85747 Garching, Germany}
\address{$^2$ GSI, D-64291 Darmstadt, Germany}
\address{$^3$ Institute of Theoretical Physics,
University of Wroclaw, PL--50204 Wroc\l aw, Poland}
\address{$^4$ Institute f\"ur Kernphysik, 
Technische Universit\"at Darmstadt,
D-64289 Darmstadt, Germany}

\begin{abstract}

We show that in the presence of spinodal instabilities which 
develop at a first order phase transition, the fluctuations of 
conserved charges can be as strong as those at the critical end 
point (CEP). In particular, the net baryon number susceptibility 
diverges as the system crosses the isothermal spinodal lines.
This indicates that charge density fluctuations can be used  
not only to probe the CEP but also the non--equilibrium first 
order chiral phase transition in heavy ion collisions. 

\end{abstract}

\pacs{25.75.Nq, 24.60.Lz}

\section{Introduction}	

The search for the critical end point (CEP)~\cite{cep} has 
attracted a considerable attention in heavy-ion phenomenology~\cite{srs}.
It is of particular interest to identify its position in the phase 
diagram and to study general properties of thermodynamic quantities 
in its vicinity. 
Modifications in the magnitude of fluctuations or the corresponding 
susceptibilities can be considered as a possible signal for
deconfinement and chiral symmetry 
restoration~\cite{qsus:lattice,srs,qsus:model}.
In this context, fluctuations related to conserved charges play an 
important role since they are directly accessible in 
experiments~\cite{fluct,fluct:qm}.

The enhancement of baryon number fluctuations could be a clear 
indication for the existence of the CEP in the QCD phase diagram. 
However, the suppression of density fluctuations along the 
first-order transition appear under the assumption that this transition 
takes place in equilibrium. This is modified when there is a 
deviation from equilibrium~\cite{our:spinodal}.
In this contribution we briefly show that enhanced baryon number 
density fluctuations is a signal for the first-order phase transition 
in the presence of spinodal decomposition.

\section{The role of spinodal instabilities in fluctuations}

In a non-equilibrium system,  a first-order 
phase transition is intimately linked with the existence of a 
convex anomaly in the thermodynamic pressure~\cite{ran}.
There is an interval of energy density or baryon number density 
where the derivative of the pressure, $\partial P/{\partial V}>0$, 
is positive. This anomalous behavior characterizes a region of 
instability in the ($T,n_q)$-plane which is bounded by the spinodal 
lines, where the pressure derivative with respect to volume vanishes. 
The derivative taken at constant temperature and that taken at 
constant entropy,
\begin{equation}
\left( \frac{\partial P}{\partial V} \right)_T=0 
\qquad{\rm and}\qquad 
\left(
\frac{\partial P}{\partial V} \right)_S=0\,,
\end{equation}
define the isothermal and isentropic spinodal lines respectively.

If the first-order phase transition takes place in 
equilibrium, there is a coexistence region, which ends at the CEP.
However, in a non-equilibrium first-order phase transition, 
the system supercools/superheats and, if driven sufficiently far 
from equilibrium, it becomes locally unstable due to the convex anomaly.
In other words, in the coexistence 
region there is a range of densities and temperatures, bounded 
by the spinodal lines, where the spatially uniform system is 
mechanically unstable.
Spinodal decomposition is thought to play a dominant role in the 
dynamics of low energy nuclear collisions in the regime of the 
first-order nuclear liquid-gas transition~\cite{ran,heiselberg}. 
In connection with the chiral and deconfinement phase transitions, 
the possibility of spinodal decomposition in heavy ion collisions
has been discussed in~\cite{ran,heiselberg,spinodal:fluc}.

\begin{figure}
\begin{center}
\includegraphics[width=7cm]{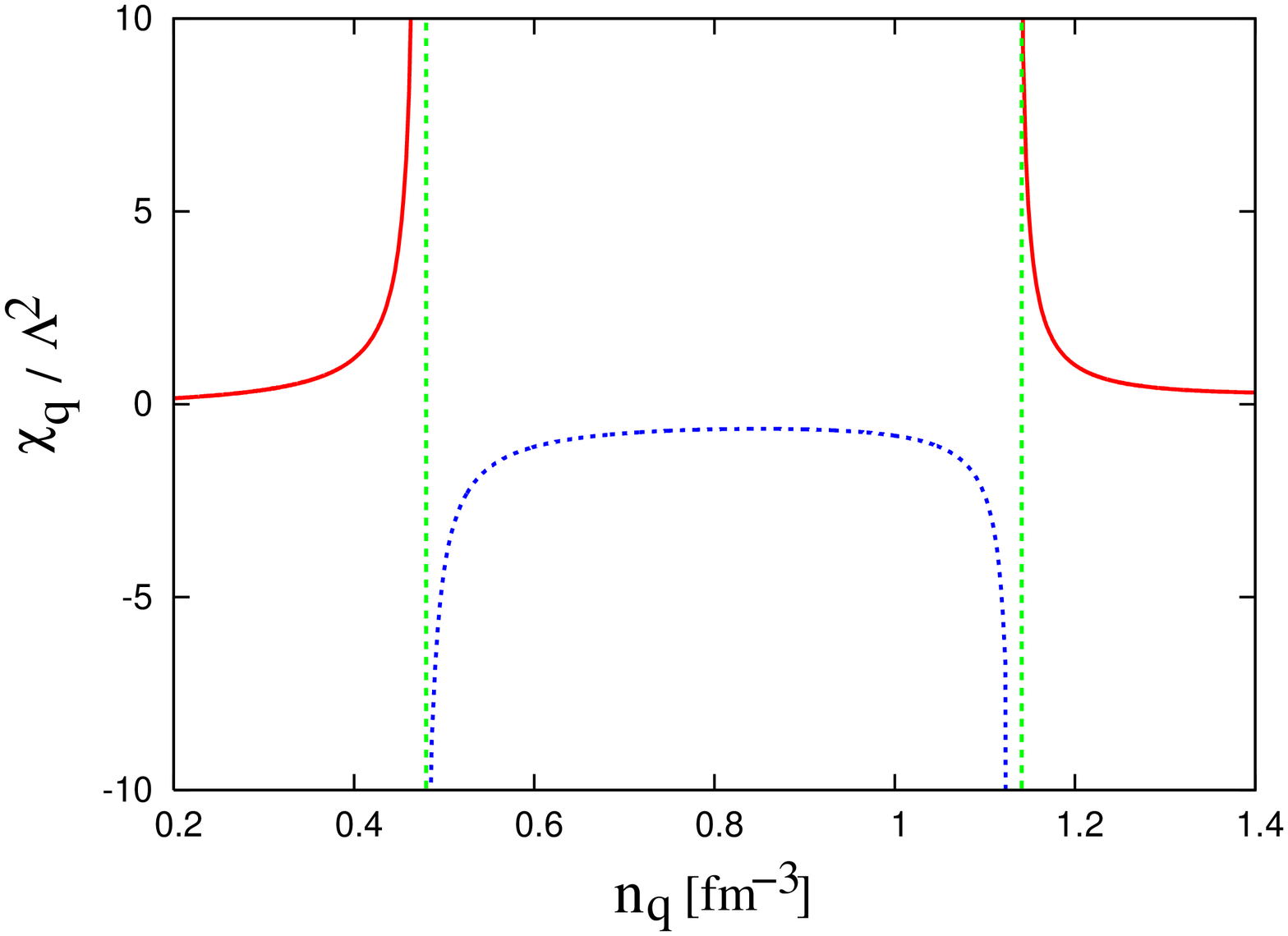}
\includegraphics[width=8cm]{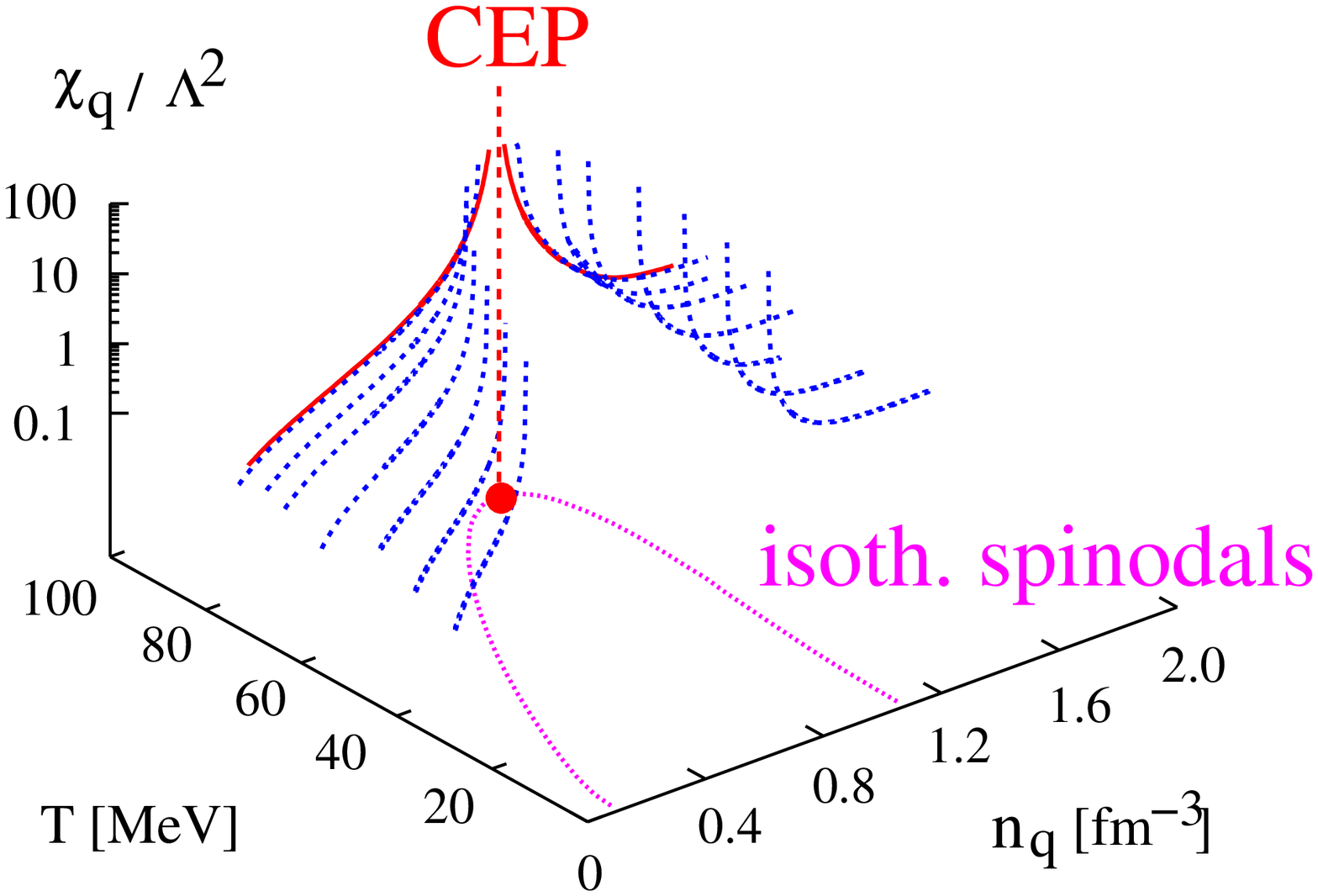}
\caption{
(Left) The net quark number susceptibility at $T=50$ MeV as a
function of the quark number density across the first-order
phase transition. (Right) The net quark number susceptibility
in the stable and meta-stable regions.
}
\label{sus_sp}
\end{center}
\end{figure}

In Fig.~\ref{sus_sp}-left we show the evolution of the net quark 
number fluctuations along a path of fixed $T=50$ MeV calculated 
in the Nambu--Jona-Lasinio (NJL) model in the mean field 
approximation~\cite{njl}.
When entering the coexistence region, a 
singularity in $\chi_q$ appears at the isothermal
spinodal lines, where the fluctuations diverge and the 
susceptibility changes sign. 
In between the spinodal lines, the susceptibility is negative. 
This implies  instabilities in the baryon number 
fluctuations when crossing from a meta-stable to an unstable phase.
The above  behavior of $\chi_q$ is a direct consequence of the 
thermodynamics relation
\begin{equation}
\left( \frac{\partial P}{\partial V} \right)_T
= - \frac{n_q^2}{V}\frac{1}{\chi_q}\,.
\label{pder}
\end{equation}
Along the isothermal spinodals the  pressure derivative  in 
Eq.~(\ref{pder}) vanishes.
Thus, for non-vanishing density $n_q$, $\chi_q$ must diverge to 
satisfy (\ref{pder}).
Furthermore, since the pressure derivative 
${\partial P}/{\partial V}|_T$ changes sign when crossing the 
spinodal line, there must be a corresponding sign change in $\chi_q$,
as seen in Fig.~\ref{sus_sp}-left. 

In Fig.~\ref{sus_sp}-right we show the evolution of the singularity
at the spinodal lines in the $T$-$n_q$ plane. The critical 
exponent at the isothermal spinodal line is found to be $\gamma=1/2$, 
with $\chi_q \sim (\mu-\mu_c)^{-\gamma}$, while $\gamma=2/3$ at the 
CEP~\cite{our:spinodal}. Thus, the singularities at the two spinodal 
lines conspire to yield a somewhat stronger divergence as they join 
at the CEP. The critical region of enhanced susceptibility around 
the CEP is fairly small~\cite{schaefer-wambach,our:njl}, 
while in the more realistic non-equilibrium system one expects 
fluctuations in a larger region of the phase diagram, i.e., 
over a broader range of beam energies, due to the spinodal instabilities.

The rate of change in entropy with respect to temperature at constant
pressure gives the specific heat expressed as
\begin{equation}
C_P
= T \left( \frac{\partial S}{\partial T} \right)_P
= TV \left[ \chi_{TT} - \frac{2 s}{n_q}\chi_{\mu T}
{}+ \left( \frac{s}{n_q} \right)^2 \chi_q \right]\,.
\end{equation}
The entropy $\chi_{TT}$ and mixed $\chi_{\mu T}$ susceptibilities 
exhibit the same critical behaviors as that of $\chi_q$ shown in 
Fig.~\ref{sus_sp}-left. Thus $C_P$ also diverges on the isothermal 
spinodal lines and becomes negative in the mixed phase~\footnote{
 The specific heat with constant volume, on the other hand,
 continuously changes with $n_q$ and has no singularities
 on the mean-field level.
}. It was  argued that in low energy nuclear collisions the negative 
specific heat could be a signal of the liquid-gas phase 
transition~\cite{chomaz}. Its occurrence has recently been reported 
as the first experimental evidence for such an anomalous
behavior~\cite{experiment}.

\section{Conclusions}

We showed that in the presence of spinodal instabilities 
the net quark number fluctuations diverge at the isothermal 
spinodal lines of the first-order chiral phase transition. 
As the system crosses this line, it becomes unstable with respect 
to spinodal decomposition. The unstable region is in principle 
reachable in non-equilibrium systems that is most likely created 
in heavy ion collisions. Consequently,  large fluctuations of 
baryon and electric charge  densities are expected to probe
not only the CEP but also a first order phase transition when the
system crosses the spinodal lines.

\section*{Acknowledgments}

C.S. acknowledges partial support by DFG cluster of excellence
``Origin and Structure of the Universe''.
K.R. acknowledges partial support of the Polish Ministry of National
Education (MENiSW) and DFG under the "Mercator program".


\vspace*{1cm}

\begin{thebibliography}{99}
\bibitem{cep}
  J.~Berges and K.~Rajagopal,
  Nucl.\ Phys.\ B {\bf 538}, 215 (1999).

\bibitem{srs}
  M.~A.~Stephanov, K.~Rajagopal and E.~V.~Shuryak,
  Phys.\ Rev.\ Lett.\  {\bf 81}, 4816 (1998);
M. Stephanov, Acta Phys.Polon. B {\bf 35}, 2939 (2004);
  Prog.\ Theor.\ Phys.\ Suppl.\  {\bf 153}, 139 (2004);
  Int.\ J.\ Mod.\ Phys.\ A {\bf 20}, 4387 (2005).

\bibitem{qsus:lattice}
S.~A.~Gottlieb {\it et al.},
Phys.\ Rev.\ Lett.\  {\bf 59}, 2247 (1987),
Phys. Rev. D {\bf 55} 6852 (1997);
R. V. Gavai and S. Gupta,
Phys. Rev. D {\bf 65} 094515 (2002);
C.~R.~Allton {\it et al.},
  Phys.\ Rev.\ D {\bf 68}, 014507 (2003);
C.~R.~Allton {\it et al.},
  Phys.\ Rev.\ D {\bf 71}, 054508 (2005);
M.~D'Elia and M.~P.~Lombardo,
Phys.\ Rev.\ D {\bf 67}, 014505 (2003);
R.V. Gavai and  S. Gupta,
Phys. Rev. D {\bf 72} 054006 (2005),
Eur. Phys. J. C {\bf 43} 31 (2005),
Phys. Rev. D {\bf 73}  014004 (2006).


\bibitem{qsus:model}
T. Kunihiro,
Phys. Lett. B {\bf 271} 395  (1991);
Y.~Hatta and T.~Ikeda,
Phys.\ Rev.\ D {\bf 67}, 014028 (2003);
  H.~Fujii,
  Phys.\ Rev.\ D {\bf 67}, 094018 (2003);
  H.~Fujii and M.~Ohtani,
  Phys.\ Rev.\ D {\bf 70}, 014016 (2004).
  P.~Costa {\it et al.},
  Phys.\ Lett.\  B {\bf 647}, 431 (2007).

\bibitem{fluct}
 S. Jeon and V. Koch,
{\it Quark Gluon Plasma 3}, Eds. R.C. Hwa and
X. N. Wang, World Scientific Publishing, 2004.

\bibitem{fluct:qm}
V.~Koch, these proceedings.

\bibitem{our:spinodal}
  C.~Sasaki, B.~Friman and K.~Redlich,
  Phys.\ Rev.\ Lett.\  {\bf 99}, 232301 (2007);
  arXiv:0709.2487 [hep-ph];
  Phys.\ Rev.\  D {\bf 77}, 034024 (2008).

\bibitem{ran}
P. Chomaz, M. Colonna and  J. Randrup,
Phys. Rept. {\bf 389},  263 (2004).

\bibitem{heiselberg} 
H.~Heiselberg, C.~J.~Pethick and D.~G.~Ravenhall,
Phys. Rev. Lett. {\bf 61}, 818 (1988);
Annals Phys.\ {\bf 223}, 37 (1993).

\bibitem{spinodal:fluc} 
D. Bower and S. Gavin, Acta Phys. Hung. A15, 269, 1219-7580 (2002);
J. Randrup, Acta Phys. Hung. A22, 69, 1219-7580 (2005);
J. Randrup, Phys. Rev. Lett. {\bf 92}, 122301 (2004);
V. Koch, A. Majumder and J. Randrup,  
Phys. Rev. C {\bf 72},  064903 (2005);
J. Polonyi, hep-th/0509078;
  K.~Paech, H.~Stoecker and A.~Dumitru,
  Phys.\ Rev.\  C {\bf 68}, 044907 (2003);
  K.~Paech and A.~Dumitru,
  Phys.\ Lett.\  B {\bf 623}, 200 (2005);
O.~Scavenius, A.~Dumitru, E.~S.~Fraga, J.~T.~Lenaghan and A.~D.~Jackson,
  Phys.\ Rev.\  D {\bf 63}, 116003 (2001);
  C.~E.~Aguiar, E.~S.~Fraga and T.~Kodama,
  J.\ Phys.\ G {\bf 32}, 179 (2006);
  E.~S.~Fraga and G.~Krein,
  Phys.\ Lett.\  B {\bf 614}, 181 (2005).

\bibitem{njl}
For a recent review, see e.g.,
M.~Buballa,
Phys.\ Rept.\  {\bf 407}, 205 (2005).

\bibitem{schaefer-wambach}
  B.~J.~Schaefer and J.~Wambach,
  Phys.\ Rev.\  D {\bf 75}, 085015 (2007);
  B.~J.~Schaefer,
  arXiv:0709.4216 [hep-ph].

\bibitem{our:njl}
  C.~Sasaki, B.~Friman and K.~Redlich,
  Phys.\ Rev.\  D {\bf 75}, 054026 (2007).

\bibitem{chomaz}
  P.~Chomaz and F.~Gulminelli,
  Nucl.\ Phys.\  A {\bf 647}, 153 (1999).

\bibitem{experiment}
  M.~D'Agostino {\it et al.},
  Phys.\ Lett.\  B {\bf 473}, 219 (2000);
  M.~Schmidt {\it et al.},
  Phys.\ Rev.\ Lett.\  {\bf 86} (2001) 1191.

\end{thebibliography}
\end{document}